\documentclass{pasj01}
\draft 
\Received{$\langle$reception date$\rangle$}
\Accepted{$\langle$acception date$\rangle$}
\Published{$\langle$publication date$\rangle$}
\begin{document}
\title{Where is the Western Part of the Galactic Center Lobe Located really?}
\author{Masato Tsuboi$^1$, Takahiro Tsutsumi$^2$,  Yoshimi Kitamura$^1$,  Ryosuke Miyawaki$^3$,  Atsushi Miyazaki$^4$, and Makoto Miyoshi$^5$ }%
\altaffiltext{1}{Institute of Space and Astronautical Science, Japan Aerospace Exploration Agency,\\
3-1-1 Yoshinodai, Chuo-ku, Sagamihara, Kanagawa 252-5210, Japan }
\email{tsuboi@vsop.isas.jaxa.jp}
\altaffiltext{2}{National Radio Astronomy Observatory,  Socorro, NM 87801-0387, USA}
\altaffiltext{3}{College of Arts and Sciences, J.F. Oberlin University, Machida, Tokyo 194-0294, Japan}
\altaffiltext{4}{Japan Space Forum, Kanda-surugadai, Chiyoda-ku,Tokyo,101-0062, Japan}
\altaffiltext{5}{National Astronomical Observatory of Japan, Mitaka, Tokyo 181-8588, Japan}

\maketitle

\begin{abstract}
The Galactic Center Lobe (GCL) is a peculiar object widely protruding from the Galactic plane toward the positive Galactic latitude, which had been found toward the Galactic Center (GC) in the early days of the radio observation.
The peculiar shape has suggested any relation with historical events, star burst, large explosion and so on in the GC. However, the issue whether the GCL is a single large structure located in the GC region is not yet settled conclusively.
In the previous observations, the silhouette against the low frequency emission was found in the western part of the GCL (WPGCL), This suggests  that the part is located in front of the GC region. On the other hand, the LSR velocity of the radio recombination line toward it was found to be as low as 0 kms$^{-1}$.  However, these observations cannot determine the exact position on the line-of-sight.  There is still another possibility that it is in the near side area of the GC region. In this analysis, we compare these results with the visual extinction map toward the GC. We found that the distribution of the visual extinction larger than 4 mag. clearly corresponds to the silhouette of the WPGCL.  The WPGCL must be located at most within a few kpc from us and not in the GC region. This would be a  giant HII region in the Galactic disk.
\end{abstract}

%%%%%%%%%%%%%%%%%%%%%%%%%%%%%%%%%%%%%%%%%%
\section{Introduction}
Peculiar objects have been observed toward the Galactic center (GC) region in wide wavelength regimes since the early days of the observation. 
Some of them are really located in the GC region and they provide the irreplaceable information to investigate the structure and activity of the region. However, the other objects are not really located in the GC region. Although the objects are ordinary objects in the regions where they are really located, the misunderstanding that they are located in the GC makes them falsely identified  as peculiar objects.

The Galactic Center Lobe (GCL) is a large scale object found in the GC region (\cite{Sofue1984}). The  peculiar shape has suggested any relation with historical events, star burst, large explosion and so on in the GC.
However, it has been also advocated that the GCL is not a single large structure in the GC region based on some circumstantial evidences. Especially, it is not yet settled conclusively where the western part of the GCL (WPGCL) is located in the line of sight (e.g.\cite{Tsuboi1986}, \cite{Law2009}, \cite{Nagoshi}).
Recently new large radio telescopes at low frequency (MWA, MeerKAT, and so on) start observing. The GC region has been often selected as their early observation targets. The peculiar shape of the GCL is imaged fascinatingly in their new maps (e.g. \cite{Heywood}, \cite{Hurley-Walker}). In these cases,  improper discussions originated from the ambiguity of the location may waste our research resources. In order to dispel this anxiety, we attempt to solve this long-standing issue making the best use of now available data including the visual extinction ($A_\mathrm{V}$) in this letter. 
Throughout this paper, we adopt 8 kpc as the distance to the GC (e.g. \cite{Boehle}). Then, $1\arcsec$ corresponds to about 0.04 pc at the distance. 

\section{Existing Observational Results of the Galactic Center Lobe}
The GCL had been observed by various telescopes since the discovery.
However, all the observational results obtained so far are not always consistent with each other. Here, we re-examine them. 
%%%%%%%%%%%%%%%%%%%%%
\begin{figure}
\begin{center}
\includegraphics[width=16cm, bb=0 0  1023 969]{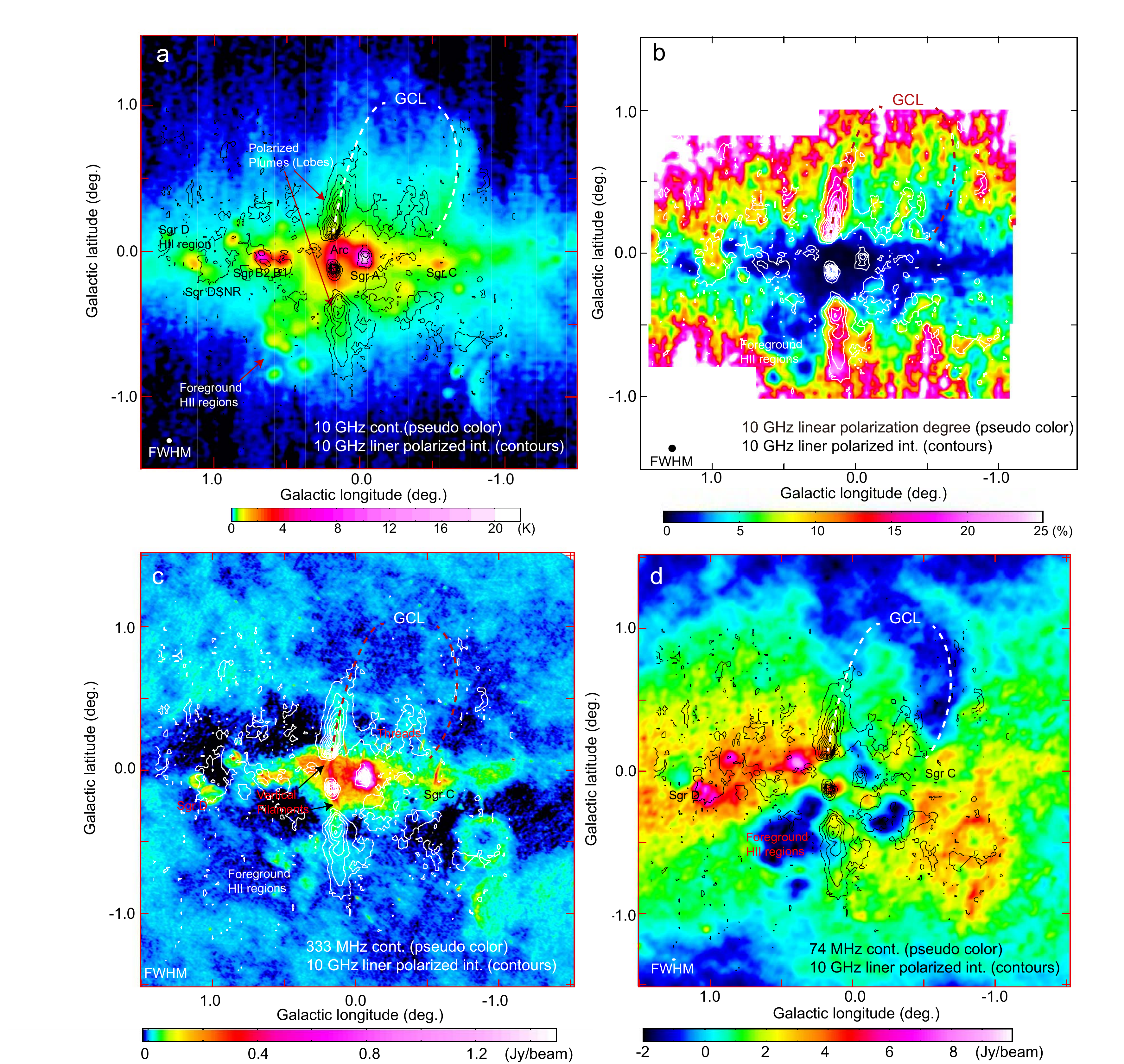}
 \end{center}
 \caption{{\bf a} $3^\circ\times 3^\circ$ continuum image at 10 GHz of the Galactic center region by the Nobeyama 45 m telescope (\cite{Handa}).  The beam size in FWHM is $3'\times3'$ shown in the bottom left corner. The contours show the linear polarized intensity at 10 GHz (\cite{Tsuboi1986}). The first contour and interval are both 50 mJy/beam (0.019 K in $T_\mathrm{B}$). The broken line curves indicate the Galactic Center Lobe. {\bf b} Image of the degree of linear polarization at 10 GHz (\cite{Tsuboi1986}). {\bf c} Continuum image at 333 MHz by VLA (\cite{LaRosa}). The beam size in FWHM is $43"\times24", PA=65^\circ$ shown in the bottom left corner. The 1 Jy/beam corresponds to 10700 K  in $T_\mathrm{B}$. {\bf d} Continuum image at 74 MHz by VLA (\cite{Brogan}). The beam size in FWHM is $114"\times60"$ shown in the bottom left corner. The 1 Jy/beam corresponds to 32600 K  in $T_\mathrm{B}$.}
 \label{Fig1}
\end{figure}
%%%%%%%%%%%%%%%%%%%%%
\subsection{Radio Continuum Images}
Figure 1a shows the $3^\circ\times 3^\circ$ continuum image at 10 GHz of the Galactic center region by 
the Nobeyama (NRO) 45 m telescope (\cite{Handa}).
The GCL had been found in the 10 GHz continuum image (\cite{Sofue1984}).
The GCL is widely protruding from the Galactic plane toward the positive Galactic latitude as an $\Omega$-shaped structure  (broken line curves). This figure also shows the linear polarized intensity at 10 GHz (contours).  
The eastern part of the GCL (EPGCL) overlaps the positive latitude part of the Polarized Plumes (or Polarized Lobes,  PPs) (\cite{Tsuboi1985}, \cite{Tsuboi1986}, \cite{Seiradakis1985}). On the other hand, the WPGCL seems to be rooted around Sgr C (also see Figure 1c). 
Figure 1b shows the image of the degree of linear polarization (DLP) at 10 GHz  (\cite{Tsuboi1986}). Although the galactic plane is not polarized except for several spots, a high DLP up to  $DLP > 25$ \% has been detected in the PPs. The EPGCL  is mainly a non-thermal structure. The degree of linear polarization in the WPGCL is very low, $DLP < 2$ \%. The low DLP may be originated by Faraday depolarization of the foreground ionized gas, or the WPGCL may be the HII region in the near side region of the GC. 

Figure 1c shows the continuum image of the Galactic center region at 333 MHz by the Karl G. Jansky Very Large Array (VLA)  (\cite{LaRosa}).  The Vertical Filaments (VFs) of the Galactic Center Arc  (\cite{Yusef-Zadeh1984}, \cite{Yusef-Zadeh1987}), threads (e.g. \cite{Morris1985}), and the linear non-thermal filaments of Sgr C  (\cite{Liszt1985}) are clearly seen in the figure. The VFs have been also observed to interact with the Galactic center molecular clouds (GCMCs) (e.g. \cite{Yusef-Zadeh1987}, \cite{Tsuboi1997}). 
The both inside ends of PPs connect to the both ends of the VFs,  which are located in the GC region or at least in the vicinity of the region. 
Although the southernmost part of the WPGCL positionally corresponds to the linear filaments of Sgr C, the WPGCL have no close connection with the other Galactic center objects including the GCMCs and so on.   The figure has negative emission areas around the strong positive emission areas because of the shortage of the short-baseline data.  
Nevertheless, the WPGCL is not identified as both a faint emission or a silhouette feature against the extended continuum emission of the GC (\cite{Rickert}).
This disappearance of the WPGCL may be caused by that the intensity difference between the WPGCL and surrounding emission decreases at this frequency. Note that the WPGCL is detected as an emission at  $\sim1.3$ GHz (\cite{Heywood}) and as a silhouette at $\sim150$ MHz (\cite{Hurley-Walker}). 

Figure 1d shows the continuum image of the Galactic center region at 74 MHz by VLA (\cite{Brogan}).
%%%%%%%%%%%%%%%%%%%%%
The WPGCL is detected as a curved silhouette feature against the continuum emission at 74 MHz. On the other hand, the EPGCL has no such feature except for a weak feature on the negative longitude side of the PPs (\cite{Nagoshi}, also see Figure 9 in \cite{Hurley-Walker}). 
Because the large part of the continuum emission at 74 MHz would come from the GC region through the non-thermal process and  the temperature of the ionized gas in the WPGCL is low compared with the brightness temperature of the background continuum emission, the silhouette feature suggests that the WPGCL is located in front of the GC region or at least on the near side of the region.
This is consistent with other silhouette features by the foreground HII regions.

%%%%%%%%%%%%%%%%%%%%%
\begin{figure}
\begin{center}
\includegraphics[width=16cm, bb=0 0  940 905]{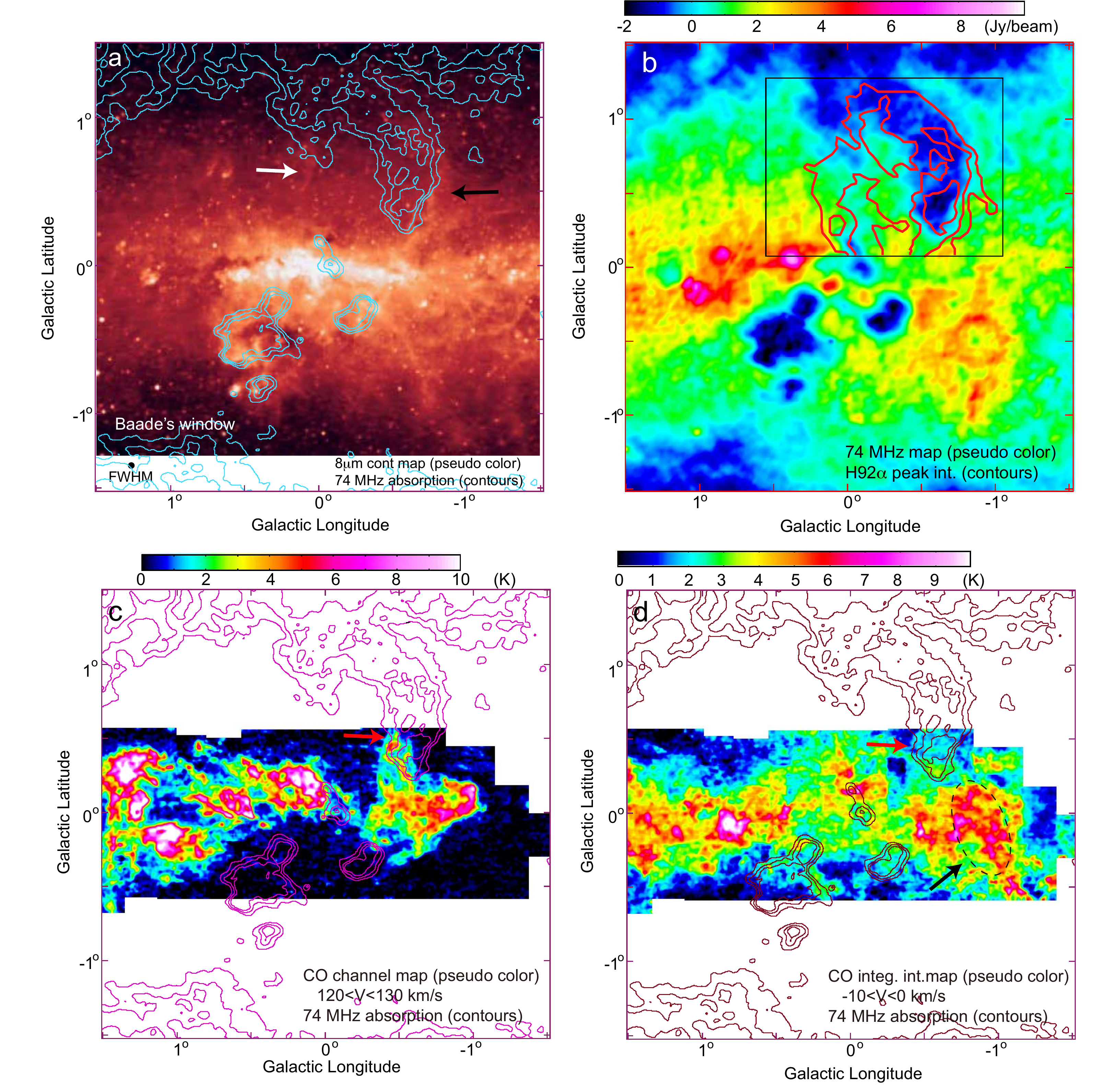}
 \end{center}
 \caption{{\bf a} $8\mu$m image of the Galactic center region by Midcourse Space Experiment (\cite{Price2001}). 
The contours show silhouette features against the continuum emission at 74 MHz. {\bf b} Continuum image at 74 MHz by VLA (\cite{Brogan}). The contours show the distribution of the H$92\alpha$ radio recombination line (RRL) (\cite{Nagoshi}). {\bf c} Integrated intensity map with the velocity range of $V_\mathrm{LSR}=120-130$ km s$^{-1}$ in the CO $J=1-0$ emission line  by the NRO 45 m telescope (\cite{Oka1998}). The contours show silhouette features against the continuum emission at 74 MHz. {\bf d} Integrated intensity map with the velocity range of $V_\mathrm{LSR}=-10-0$ km s$^{-1}$ in the CO $J=1-0$ emission line. The contours show silhouette features against the continuum emission at 74 MHz. }
 \label{Fig2}
\end{figure}
%%%%%%%%%%%%%%%%%%%%%

\subsection{Ionized and Molecular Gas Images}
Figure 2a shows the image at $8.3\mu$m by Midcourse Space Experiment (MSX) (\cite{Price2001}). The  contours indicate the silhouette features at 74 MHz (see Figure 1d). The IR counterpart of the WPGCL can be identified (black arrow).  This also corresponds to AFGL5376, which is around $l\sim -0.^{\circ}5, b\sim0.^{\circ}5$ in IR maps. Meanwhile the EPGCL has no clear counter part although the ``Double Helix Nebula"  is identified (white arrow; \cite{Morris2006}).

Figure 2b shows the comparison between the silhouette feature of the WPGCL at 74 MHz and the distribution of the H$92\alpha$ radio recombination line (RRL) (\cite{Nagoshi}). 
The LSR velocity of the H$92\alpha$ RRL is measured to be $V_\mathrm{LSR}\sim-4-10$ km s$^{-1}$ (also see  \cite{Law2009}). The distribution of the RRL clearly corresponds to the silhouette feature (\cite{Nagoshi}). 
The velocity of $\sim0$ km s$^{-1}$ in the WPGCL allows the possibility that it is in the area intervening between the GC and us, for example, foreground spiral arms.   
In addition, the LSR velocity of the H$70\alpha$ RRL toward the Sgr C HII region has been observed to be $V_\mathrm{LSR}\sim-65$ km s$^{-1}$ (\cite{Liszt1995}). The Sgr C HII region is observed to be associated with the GCMCs with around this velocity (e.g. \cite{Tsuboi1991}, \cite{Liszt1995}). The velocity is different from that of the RRL toward the WPGCL as mentioned above. This suggests that there is no physical connection between the WPGCL and Sgr C.

Figure 2c shows the integrated intensity map with the LSR velocity range of $V_\mathrm{LSR}=120-130$ km s$^{-1}$ in the CO $J=1-0$ emission line  by the NRO 45 m telescope (\cite{Oka1998}). The contours show the silhouette feature at 74 MHz. 
There is the molecular cloud positionally corresponding to the southernmost part of the WPGCL or AFGL5376 (red arrow in Figure 2c).  The molecular cloud is identified in the velocity range of $V_\mathrm{LSR}\sim110-140$ km s$^{-1}$ (\cite{Uchida}, also see \cite{Oka1998}). 
The large positive velocity suggests that the molecular cloud is located in the GC region. However, the velocity is different from the velocity of the RRL toward the WPGCL, $V_\mathrm{LSR}\sim-4-10$ km s$^{-1}$ (\cite{Nagoshi}). This suggests that there is no physical connection between the WPGCL and the molecular cloud.
On the other hand, Figure 2d shows the integrated intensity  map with the velocity range of $V_\mathrm{LSR}=-10 - 0$ km s$^{-1}$.   A faint molecular cloud  in this channel map (red arrow) would exist around $l\sim -0.^{\circ}5, b\sim0.^{\circ}5$ or the southernmost part of the WPGCL. The velocity of the molecular cloud is consistent with the velocity of the ionized gas in the WPGCL.  The molecular cloud is extended up to $b\sim0.^\circ 8$ (see Figure 4 in \cite{Takeuchi2010}). This is thought to be the counterpart of the WPGCL. 
Another molecular cloud is located around $l=-0.^{\circ}8, b=-0.^{\circ}2$ (broken line oval).  This may be the negative latitude extension of the WPGCL.

%%%%%%%%%%%%%%%%%%%%%
\begin{figure}
\begin{center}
\includegraphics[width=16cm, bb=0 0  874 445]{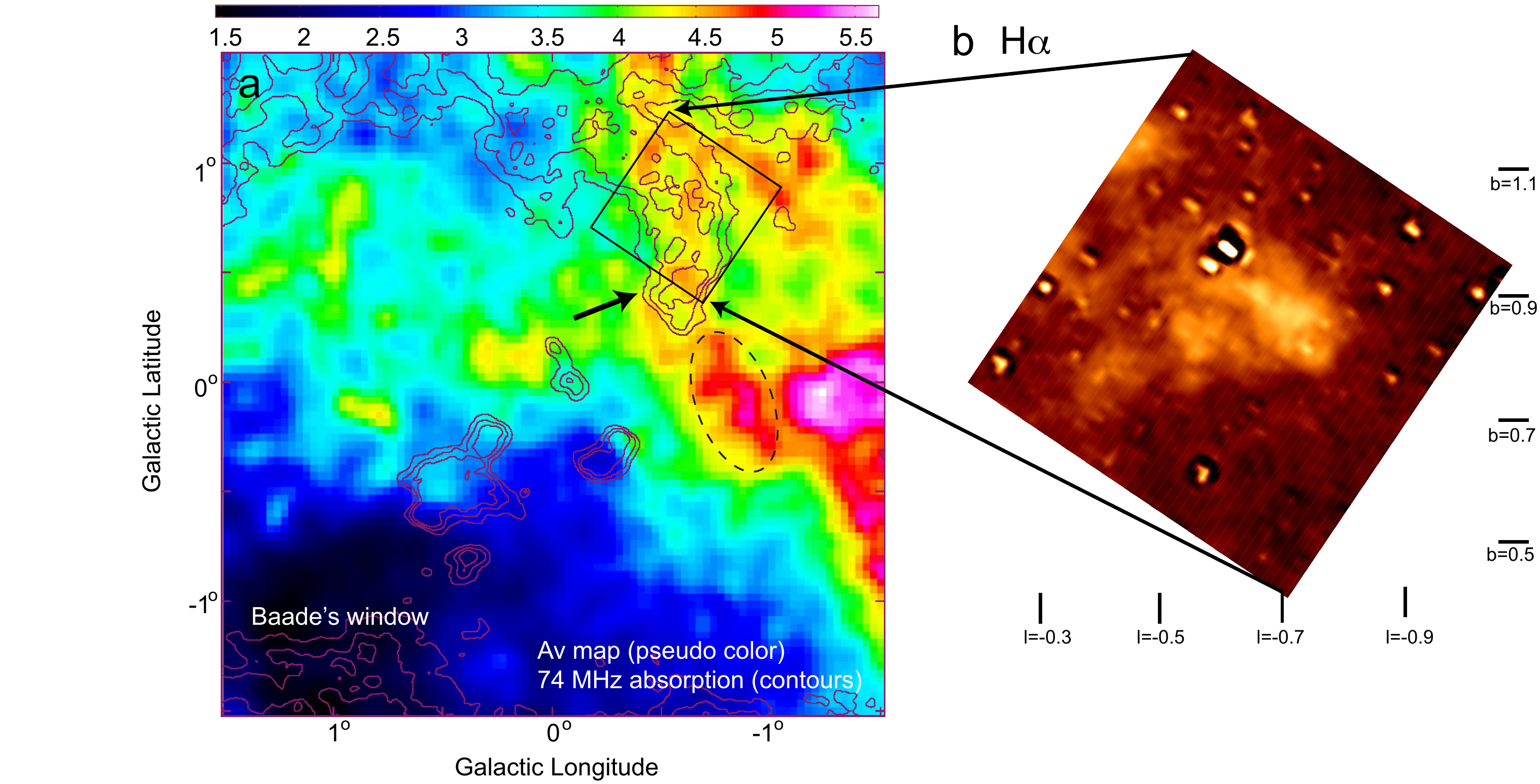}
 \end{center}
 \caption{{\bf a} Image of visual extinction, $A_\mathrm{V}$, in the GC region based on ``Degitized Sky Survey" (pseudocolor, \cite{Dobashi}). The contours show silhouette features against the continuum emission at 74 MHz (\cite{Brogan}).  {\bf b} Map of the western part of the GCL in the H$\alpha$ emission (\cite{Gaustad}). The map area is shown as a rectangle in Figure 3a. }
 \label{Fig3}
\end{figure}
%%%%%%%%%%%%%%%%%%%%%

\section{Comparison with $A_\mathrm{V}$ Map}
Figure 3a shows the $A_\mathrm{V}$ map of the GC region based on ``Degitized Sky Survey(DSS)" (pseudocolor, \cite{Dobashi}).
The contours show the silhouette features at 74 MHz. There is an $A_\mathrm{V}$ feature (black arrow) clearly corresponding to the silhouette of the WPGCL.  The $A_\mathrm{V}$ feature and silhouette must be the different aspects of the same object.
The DSS is the survey observation using visible light. The containing stars are located at most within a few kpc from us toward the GC because the farther stars are not seen by strong visual extinction near the galactic plane.
Therefore the object which makes the $A_\mathrm{V}$ feature must be located at most within a few kpc.  That is, the WPGCL is not located in the nearside region of the GC because the distance to the GC is 8 kpc. 
%The faint trace of this $A_\mathrm{V}$ feature is also seen in the $A_\mathrm{K}$ map (Figure 1 in \cite{Dutra}).
In addition, another $A_\mathrm{V}$ feature connecting the WPGCL is located around $l=-0.^{\circ}8, b=-0.^{\circ}2$ (broken line oval). This feature is though to be the CO negative latitude extension of the WPGCL as mentioned in the previous section.

The Hydrogen column density is derived from the visual extinction, $A_\mathrm{V}$, using the following formula (e.g. \cite{Guver}),
\begin{equation}
\label{ }
N_\mathrm{H} \mathrm{[cm^{-2}]}= 2.2\times10^{21}A_\mathrm{V}\mathrm{[mag.]}.
\end{equation}
The mean visual extinction of the $A_\mathrm{V}$ feature is  measured to be $\bar{A_\mathrm{V}} \sim4.5$ mag. (see Figure 3).
The mean Hydrogen column density is estimated to be $\bar{N_\mathrm{H}} = 1\times10^{22}$ cm$^{-2}$. This is considered to be the mean Hydrogen molecule column density of $\bar{N_\mathrm{H2}} \sim 5\times10^{21}$ cm$^{-2}$. On the other hand, the Hydrogen molecule column density is also derived from the CO integrated intensity using the following formula (e.g. \cite{Bolatto}),
\begin{equation}
\label{ }
N_\mathrm{H2} \mathrm{[cm^{-2}]}= 2\times10^{20}\int T_\mathrm{MB}dv \mathrm{[K~km s^{-1}]}.
\end{equation}
The mean CO integrated intensity of the $A_\mathrm{V}$ feature is measured to be  $\int T_\mathrm{MB}dv \sim22$ K km s$^{-1}$ (see Figure 2d). 
The mean Hydrogen molecule column density is estimated to be $\bar{N_\mathrm{H2}} \sim 4.4\times10^{21}$ cm$^{-2}$. This value is consistent with that from $A_\mathrm{V}$. 
Assuming that the feature is located at $D=2$ kpc and the path length is the same as the width of the silhouette, the path length is $d=7$ pc.
The Hydrogen molecule density is estimated to be $\bar{n_\mathrm{H2}}\sim\frac{\bar{N_\mathrm{H2}}}{d}=200$ cm$^{-3}$. This value is higher than the ``effective" critical density of the CO $J=1-0$ emission line, which is estimated to be $n_\mathrm{H2, CO, crit.}\sim100$ using the RADEX simulation (\cite{VanderTak}), but lower than that of the CS emission lines ($n_\mathrm{H2, CS, crit.}\sim10^4$). This is consistent with that the feature is not identified in the channel maps of the CS emission lines (e.g. $J=2-1$; \cite{Bally1987},  $J=1-0$; \cite{Tsuboi1999}). If the feature is located at $D=8$ kpc or in the GC, the Hydrogen molecule density is estimated to be $\bar{n_\mathrm{H2}}\sim50$ cm$^{-3}$. This value is lower than the ``effective" critical density of the CO $J=1-0$ emission line. The detection in the CO $J=1-0$ emission line would be inconsistent with the Hydrogen molecule density.

The  electron density, $n_{\mathrm e}$, in the WPGCL is estimated from the continuum brightness temperature of $T_{\mathrm B}\sim0.3$ K at 10 GHz (\cite{Handa}),  the electron temperature of $T^\ast_{\mathrm e}=4200$ K (4400 K;\cite{Nagoshi}, 4000 K;\cite{Law2009})  and the path length of  $d=7$ pc using the well-known formula which is given by 
\begin{equation}
\label{ } 
n_\mathrm{e}[\mathrm{cm}^{-3}]=
\bigg\{\frac{[T_{\mathrm B}/\mathrm{K}][T_\mathrm{e}/\mathrm{K}]^{0.35}[\nu/\mathrm{GHz}]^{2.1}}{8.24\times10^{-2}a(\nu,T_\mathrm{e})[d/\mathrm{pc}]}\bigg\}^{0.5},
\end{equation}
(\cite{Altenhoff}). The correction factor is assumed to be $a=1$. The electron density is derived to be $n_\mathrm{e}\sim $ 35 cm$^{-3}$.  The electron density and physical size ($\sim7\mathrm{pc}\times \sim35\mathrm{pc}$) of the WPGCL are consistent with typical values of giant HII regions in the Galactic disk. Figure 3b shows the map of the WPGCL in the H$\alpha$ emission line ($656.28$ nm) (\cite{Gaustad}). The map area is shown as a rectangle in Figure 3a. We identify an H$\alpha$ emission feature correlated with the northern half of the low frequency absorption ($b\gtrsim0.8^\circ$) in the map. Because the H$\alpha$ emission line is in visible light, this detection also indicates that the WPGCL is located in the region much nearer than the GC.

%\section{Summary}
\begin{ack}  
This work is supported in part by the Grant-in-Aids from the Ministry of Eduction, Sports, Science and Technology (MEXT) of Japan, No.16K05308 and No.19K03939. 
We are grateful to Dr. C. L. Brogan at NRAO for providing the FITS file of the continuum data by VLA at 74 MHz.  
This paper is partly based on the $A_\mathrm{v}$ database by Prof. K. Dobashi at Tokyo Gakugei University. We thank him very much for providing the FITS file.  This paper is partly based on the Southern H-Alpha Sky Survey Atlas.

\end{ack}

\end{document}